\documentclass[prd,twocolumn,showpcs,amsmath,amssymb,nofootinbib,preprintnumbers,balancelastpage]{revtex4}
\pdfoutput=1
\usepackage{epsfig}
\usepackage{amsmath}
\usepackage{bm}
\usepackage{times}
\usepackage{graphicx}
\usepackage{color}
\usepackage{slashed}
\usepackage{graphicx}
\usepackage{amsmath}
\usepackage{tikz}
\usetikzlibrary{positioning,shapes}
\usepackage{relsize,caption} 
\usepackage[latin1]{inputenc}
\usepackage{tikz}
\usetikzlibrary{trees}
\usetikzlibrary{decorations.pathmorphing}
\usetikzlibrary{decorations.markings}
\usetikzlibrary{positioning,arrows}
\usetikzlibrary{decorations.pathmorphing}
\usetikzlibrary{decorations.markings}
\usetikzlibrary{decorations.pathreplacing,calc}
\usetikzlibrary{decorations.pathmorphing,decorations.markings,trees,positioning,arrows}   
\newif\ifmirrorsemicircle

\def\bea{\begin{eqnarray}}
\def\eea{\end{eqnarray}}
\def\bean{\begin{equation*}}
\def\eean{\end{equation*}}

\begin{document}

\preprint{UCI-HEP-TR-2015-05}

\title{Baryon Number as the Fourth Color}

\author{Bartosz~Fornal}
\affiliation{\\ Department of Physics and Astronomy, University of California, Irvine, CA 92697, USA}
\author{Arvind~Rajaraman}
\affiliation{\\ Department of Physics and Astronomy, University of California, Irvine, CA 92697, USA}
\author{Tim~M.~P.~Tait}
\affiliation{\\ Department of Physics and Astronomy, University of California, Irvine, CA 92697, USA}
\date{\today}

\begin{abstract}
We propose an extension of the Standard Model in which baryon number is promoted to be part of a non-Abelian gauge symmetry at high energies. 
Specifically, we consider the gauge group $SU(4) \times SU(2)_L \times U(1)_X$, 
where the $SU(4)$ unifies baryon number and color. This symmetry is spontaneously broken down to the 
Standard Model gauge group at a scale which can be as low as a few TeV. 
The $SU(4)$ structure implies that each SM quark comes along with an uncolored quark partner, the lightest of which is stabilized by the
generalized baryon number symmetry
and can play the role of dark matter.
We explore circumstances under which one can realize a model of asymmetric dark matter 
whose relic abundance is connected to the observed baryon asymmetry, and 
discuss unique signatures that can be searched for at the LHC. 
\vspace{11mm}
\end{abstract}

\maketitle

\section{Introduction}

The Standard Model (SM) provides an extremely successful description of the world. However, despite its indisputable virtues, 
many facts hint that it is just a low-energy description of a more fundamental theory. 

In particular, the SM by itself does not contain a good dark matter candidate within its spectrum, nor does it offer an acceptable explanation for the observed 
baryon asymmetry in the universe. 
Although there exists a possibility that dark matter might be a secluded particle only connected to the SM  via gravity, it remains certain that a successful 
baryogenesis mechanism will require an extension of the SM, at least at high energies.  
Indeed, one of the peculiar mysteries of baryogenesis is the fact that it involves an interaction which violates conservation of baryon number \cite{Sakharov:1967dj},
and yet searches for proton decay require that the interactions mediating proton decay be extraordinarily suppressed.
Given the observational evidence that the energy densities of dark matter and of baryons in the Universe are similar,
it is tempting to hypothesize that the two mysteries may be related in some way.  Models of {\em asymmetric dark matter} attempt to
engineer this connection by hypothesizing that the dark matter has its own particle-antiparticle asymmetry which is connected
in some way to the one
observed in 
baryons \cite{Nussinov:1985xr,Kaplan:1991ah,Hooper:2004dc,Kaplan:2009ag,Petraki:2013wwa,Zurek:2013wia}.

In this paper we construct a model of asymmetric dark matter which arises as a by-product of promoting baryon number (together with color) into a
non-Abelian gauge symmetry.
The SM gauge group is extended to $SU(4) \times SU(2)_L \times U(1)_X$, in which color is unified with the gauged baryon number symmetry into 
the single gauge group $SU(4)$. The model contains a viable dark matter candidate which is the fourth component of one of the $SU(4)$ quadruplets also containing the
SM quarks.  We explore the possibility of a successful mechanism for baryogenesis and discuss the properties of the dark matter candidate, 
whose mass is fixed through the connection between the dark matter relic density and the baryon asymmetry.  
We also note that the nonrenormalizable proton decay operators are at dimension seven, so the model does not require a huge desert above the electroweak scale.

The layout of the paper is the following: In Section II we present the particle content of the theory, list the possible interactions and discuss the symmetry breaking pattern. In Section III we show how the baryogenesis mechanism works in our model and analyze the dark matter candidate. In Section IV we discuss collider signatures and constraints on the model.  The conclusions are presented in Section V.

\section{Model}
\noindent
Our model is built on an extension of the SM gauge group:
\begin{eqnarray*}
&&SU(4) \times SU(2)_L \times U(1)_X \ ,
\end{eqnarray*}
where $X$ is a linear combination of hypercharge and the diagonal generator of $SU(4)$, and $SU(3)_C$ is contained as a subgroup
of $SU(4)$.
The SM quarks of each family are promoted to quadruplets of $SU(4)$,
\bea
&& \hspace{-10mm}\hat{Q}_{i L} \equiv \left(\!
  \begin{array}{c}
    Q^r_i \\
    Q^b_i \\
    Q^g_i \\
    \tilde{Q}_i \\
  \end{array}\!
\right)_{\!\! L} \!\! , \ \ \ \ \hat{u}_{R} \equiv \left(\!
  \begin{array}{c}
    u^r \\
    u^b \\
    u^g \\
    \tilde{u} \\
  \end{array}\!
\right)_{\!\!R} \!\! , \ \ \ \ \hat{d}_{R} \equiv \left(\!
  \begin{array}{c}
    d^r \\
    d^b \\
    d^g \\
    \tilde{d} \\
  \end{array}\!
\right)_{\!\!R} \!\! ,
\eea
where $rbg$ are the values of the color index and the
$i$ are the $SU(2)$ indices on the quark doublet.  We have suppressed a generation index running from 1 to 3 on each field.  As a consequence of
embedding $SU(3)_C$ into $SU(4)$, each quark acquires a color singlet ``quark partner", denoted by $\tilde{Q}_i$, $\tilde{u}$, and $\tilde{d}$, 
respectively.\footnote{A model based on $SU(4)$ with the quark partners identified as the SM leptons 
(as occurs in Pati-Salam unification \cite{Pati:1974yy}) was
proposed in Ref.~\cite{Perez:2013osa}.}
The SM leptons remain relatively unchanged (with $X=Y$) and, in addition, we introduce $SU(4)$ singlet fields $Q^\prime_{i R}$, $u^\prime_L$, 
and $d^\prime_L$, which after $SU(4)$ breaking will marry the $\tilde{Q}_{iL}$, $\tilde{u}_R$, and $\tilde{d}_R$:
\bea
 \hspace{4mm}Q^\prime_{i R} \ , \ \ u^\prime_L \ , \ \ d^\prime_L \ , \ \ l_{i L}  \ ,  \ \ e_R \ .
\eea
From here on we will suppress the $SU(2)$ indices.  
This simple matter content successfully includes the SM fermions, and is anomaly free.  As we
will see below, it also contains an electrically neutral Dirac fermion which can play the role of dark matter.  The scalar sector
contains an $SU(4)$ quadruplet Higgs $\hat{\Phi}$ to break $SU(4)$ and the $SU(2)_L$ doublet Higgs $H$ to break the electroweak symmetry.
The scalar and fermion field content and their gauge representations are summarized
in Table~\ref{table2}.
{\renewcommand{\arraystretch}{1.4}\begin{table}[t!]
\begin{center}
    \begin{tabular}{| c || c | c | c |}
    \hline
       \ \ \ \ \ field \ \ \ \ \  & \raisebox{0ex}[0pt]{$ \ \ \ SU(4) \ \ \ $} & \raisebox{0ex}[0pt]{$ \ \ \ SU(2)_L \ \ \ $} & \raisebox{0ex}[0pt]{$ \ \ \ U(1)_X \ \ \  $}  \\ \hline\hline
         $\hat{Q}_{L}$ & $4$ & $2$ & $0$  \\ \hline
         $\hat{u}_R$ & $4$ & $1$ &  $1/2$  \\ \hline
          $\hat{d}_R$ & $4$ & $1$ &  $-1/2$  \\ \hline
               $Q'_{R}$ & $1$ & $2$ & $-1/2$  \\ \hline
         $u'_L$ & $1$ & $1$ &  $0$  \\ \hline
          $d'_L$ & $1$ & $1$ &  $-1$  \\ \hline
           $l_L$ & $1$ & $2$ &  $-1/2$  \\ \hline
            $e_R$ & $1$ & $1$ &  $-1$  \\ \hline
                    $H$ & $1$ & $2$ &  $1/2$  \\ \hline
         $\hat{\Phi}$ & $4$ & $1$ &  $1/2$  \\ \hline
    \end{tabular}
\end{center}
\caption{{Minimal particle content of an anomaly free extension of the Standard Model with color and  baryon number unified into $SU(4)$.}}
\label{table2}
\end{table}}
The Lagrangian density can be written as:
\bea
\mathcal{L} \ =\  \mathcal{L}_{\rm gauge}  +  \mathcal{L}_{\rm kin}  + \mathcal{L}_{\rm Higgs} - \mathcal{L}_{\rm Y1} - \mathcal{L}_{\rm Y2} - \mathcal{L}_{\rm Y3} \ ,
\eea
with terms:
\vspace{1mm}
\bea
\mathcal{L}_{\rm gauge}&=& -\tfrac{1}{4} G^A_{\mu\nu} G^{A \, \mu\nu}  -\tfrac{1}{4} W^a_{\mu\nu} W^{a \, \mu\nu}  -\tfrac{1}{4} X_{\mu\nu} X^{\mu\nu} \ ,\nonumber\\ \nonumber\\[-5pt]
\mathcal{L}_{\rm kin}&=& \ \bar{\hat{Q}}_{L} \,i\slashed{D}\, \hat{Q}_{L} + \bar{\hat{u}}_{R} \,i\slashed{D}\, \hat{u}_{R} + \bar{\hat{d}}_{R} \,i\slashed{D}\, \hat{d}_{R} \nonumber\\
&&+ \ \bar{l}_{L} \,i\slashed{D}\, l_{L} + \bar{e}_{R} \,i\slashed{D}\, e_{R} \ \nonumber\\
&&+ \ \bar{Q}'_{R} \,i\slashed{D}\, Q'_{R} + \bar{u}'_{L} \,i\slashed{D}\, u'_{L} + \bar{d}'_{L} \,i\slashed{D}\, d'_{L}  \ , \nonumber\\   \nonumber \\[-5pt]
\mathcal{L}_{\rm Higgs} &=&  |D_\mu H|^2 + \ |D_\mu \hat{\Phi}|^2 + \mu^2 |H|^2 - \tfrac{1}{2} \lambda |H|^4 \nonumber\\
&&  + \ \mu_4^2 |\hat{\Phi}|^2 - \tfrac{1}{2}\lambda_4 | \hat{\Phi}|^4 
- \lambda_2 |H|^2 | \hat{\Phi}|^2,\nonumber\\   \nonumber \\[-5pt]
 \mathcal{L}_{\rm Y1} &=& y_u^{ab}\, \bar{\hat{Q}}_L^a\, \tilde{H} \,\hat{u}_R^b + y_d^{ab}\, \bar{\hat{Q}}_L^a \,H \,\hat{d}_R^b \nonumber\\   
 && + \ y_e^{ab}\, \bar{l}^a_L\,H \,e_R^b + {\rm h.c.} ,\nonumber\\    \nonumber\\[-5pt]
 \mathcal{L}_{\rm Y2} &=& {{y'}}_{\!\!u}^{ab}\, \bar{Q'}_{\!\!R}^a\,\tilde{H} \, {u'}_{\!\!L}^b + {y'}_{\!\!d}^{ab}\, \bar{Q'}_{\!\!R}^a \,H \,{d'}_{\!\!L}^b + {\rm h.c.} , \nonumber\\  \nonumber\\[-5pt]
 \mathcal{L}_{\rm Y3} &=&
 Y_Q^{ab}\,\bar{\hat{Q}}_L^a \, \hat{\Phi}\,  {Q'}_{\!\!R}^{b}+ Y_u^{ab}\,\bar{\hat{u}}_R^a \, \hat{\Phi}\, {u'}_{\!\!L}^{b}\nonumber\\
  &&  + \ Y_d^{ab} \, \bar{\hat{d}}_R^a \, \hat{\Phi}\, {d'}_{\!\!L}^{b}+ {\rm h.c.} \ .
  \label{lagr}
\eea
The gauge covariant derivative has the form:
\bea
D_\mu = \partial_\mu + i g_4 \hspace{0.4mm}G_{\mu}^A T^A + i g_2 W_\mu^a t^a + i g_X X_\mu X \ ,
\eea
where $T^A$, $t^a$, and $X$ are the $SU(4)$, $SU(2)_L$, and $U(1)_X$ generators, respectively.
The $y^{ab}$ are the SM Yukawa matrices, with generation indices $a, b = 1, 2, 3$, giving masses to the SM fields and 
quark partners through electroweak symmetry breaking. 
The $Y^{ab}$ and $y^{\prime ab}$ are Yukawa couplings generating masses only for the quark partner fields through $SU(4)$ and
$SU(2)$ spontaneous symmetry breaking, respectively.  It is generically true for phenomenologically viable parameters
that $Y \langle \hat{\Phi} \rangle \gg y \langle H \rangle$ and  $Y \langle \hat{\Phi} \rangle \gg y^\prime \langle H \rangle$, so the quark partners are to good approximation vector-like
under the electroweak symmetry.

\subsection{Gauge Bosons}

The potential for the quadruplet Higgs induces a vacuum expectation value (VEV)
\bea
\langle\hat\Phi\rangle = \frac{1}{\sqrt2}\left(0 \ \ 0 \ \  0 \ \ V\right)^T ,
\eea
responsible for the breaking:
\bea
SU(4)\!\times \!SU(2)_L \!\times\! U(1)_X \xrightarrow{\langle\hat\Phi\rangle} SU(3)_C \!\times\! SU(2)_L \!\times \!U(1)_Y \nonumber ,
\eea
at which point we have the SM.  In addition, 
the SM Higgs $H$, a doublet under $SU(2)_L$ and with $X$ charge $1/2$, breaks
$SU(2)_L \times U(1)_Y \rightarrow U(1)_{\rm EM}$ as usual.

After the $SU(4)$ breaks, the vector bosons corresponding to its 15 generators, $T^A$, $A = 1, ..., 15$, can be divided into three groups.
$G^{1...8}_\mu$ remain massless with residual $SU(3)_C$ symmetry and play the role of the SM gluons.  The six $G^{9-14}_\mu$ organize into
thee complex vector fields $G^\alpha_\mu$ transforming as a triplet under $SU(3)_C$ with masses
\bea
m_{G} = \frac{1}{2} \, g_4 \, V
\eea
and
mediate interactions involving one ordinary quark together with its uncolored partner.  The last gauge boson, corresponding to the generator
$T^{15} = {\rm diag}(1, 1, 1, -3)/ (2\sqrt6)$ (which generates baryon number), mixes with the $U(1)_X$ boson
to produce a massless state corresponding to the SM hypercharge
interaction,
\bea
Y & = & X + \sqrt{\frac{2}{3}} ~ T^{15}  =  X +  \frac{1}{6} \left(
                      \begin{array}{cccc}
                        1 & 0 & 0 & 0 \\
                        0 & 1 & 0 & 0 \\
                        0 & 0 & 1 & 0 \\
                        0 & 0 & 0 & -3 \\
                      \end{array}
                    \right),
                    \label{111-3}
\eea
 and a massive state $Z^\prime_\mu$,
 \bea
m_{Z^\prime} = \frac{1}{2}\sqrt{g_X^2+ \tfrac{3}{2} g_4^2} \ \ V \ ,
\eea
 which couples to pairs of quarks, quark partners, and leptons.
 
 It is convenient to reparametrize the gauge couplings in terms of the hypercharge coupling $g_Y$ and
 a mixing angle $\theta_4$,
 \bea
g_Y  =  \frac{g_X g_4}{\sqrt{\frac{2}{3}g_X^2+  g_4^2}} \ , \ \ \  \ \ \ 
\sin{\theta_4}  \equiv  \frac{g_X}{\sqrt{g_X^2+ \frac{3}{2} g_4^2}} \ .
%\cos{\theta_4} \equiv \frac{ g_4}{\sqrt{\frac{2}{3}g_X^2+ g_4^2}}  \ .
\eea
Note that at the scale $V$, the $SU(4)$ coupling $g_4$ matches onto the strong coupling constant $g_3$,
and thus the measured value of $g_Y$ determines $\theta_4$ for a given $V$.  For $V \sim$~TeV,
$\sin \theta_4 \approx 0.28$.
The mass eigenstates are:
\bea
\left(\!\!
  \begin{array}{c}
    {Z^\prime_\mu} \\
    B_\mu \\
  \end{array}\!\!
\right)
= \left(\!
  \begin{array}{cc}
    \cos{\theta_4} & -\sin{\theta_4}\\
   \sin{\theta_4} & \cos{\theta_4} \\
  \end{array}\!
\right)
\left(\!\!
  \begin{array}{c}
    {G^{15}_\mu} \\
    X_\mu \\
  \end{array}\!\!
\right) ,
\label{Zprime}
\eea 
and 
the coupling of the $Z^\prime$ to a pair of fermions with $T^{15}$ and hypercharge $Y$ can be written as,
\bea
\frac{g_Y}{\sin \theta_4 \cos \theta_4}
\left( -\sqrt{\frac{2}{3}} T^{15} + Y \sin^2 \theta_4 \right) .
\label{Zcouplings}
\eea

\subsection{Quark Partners}

The $SU(4)$ structure demands that each SM quark should come along with an uncolored partner field.  
After $SU(4)$ breaking, the Yukawa interactions $Y_Q$, $Y_u$, and $Y_d$ partner these fields
with the $Q^\prime$, $u^\prime$ and $d^\prime$ to form vector-like fermions under the electroweak
interaction.

After electroweak symmetry breaking, the SM Higgs mixes the elements of the quark partner
doublets with the singlets.  Writing the elements of the $SU(2)$ doublet
as $\tilde{Q} = (\tilde{U}, \tilde{D})$, the mass terms look like:
\bea
& & \hspace*{-0.75cm}
\frac{1}{\sqrt{2}}
\left(\!
\overline{\tilde{U}}_{\!L} ~~ \overline{u}^\prime_L
\right)
 \left(\!
  \begin{array}{cc}
    Y_Q V & y_u v \\
    \left( y^\prime_u v \right)^\dagger & \left( Y_u V \right)^\dagger \\
  \end{array}\!
\right)
\left(\!\!
  \begin{array}{c}
   U^\prime_R \\
    \tilde{u}_R \\
  \end{array}\!\!
\right) \nonumber \\
& & \hspace*{-0.75cm} + \frac{1}{\sqrt{2}}
\left(\!
\overline{\tilde{D}}_L ~~ \overline{d}^\prime_L
\right)
 \left(\!
  \begin{array}{cc}
    Y_Q V & y_d v \\
    \left( y^\prime_d v \right)^\dagger & \left( Y_d V \right)^\dagger \\
  \end{array}\!
\right)
\left(\!\!
  \begin{array}{c}
   D^\prime_R \\
    \tilde{d}_R \\
  \end{array}\!\!
\right)
+ {\rm h.c.} \ ,
\eea
where the $y$, $y^\prime$, and $Y$ are $3 \times 3$ matrices in flavor space
and $v \simeq 246$~GeV is the SM Higgs VEV.  In general, these are diagonalized by four $6 \times 6$ unitary matrices,
producing six Dirac mass eigenstates which are electrically neutral combinations of $\tilde{u}$ and $\tilde{U}$ (the lightest of which, say $\tilde{u}'_1$,
can be stable and play the role of dark matter) and six electric charge $\pm 1$ combinations of $\tilde{d}$
and $\tilde{D}$.  Couplings to the $W$ and $Z$ bosons are controlled by the admixture of doublet and singlet in each mass eigenstate.

As with any theory which extends the notion of flavor beyond the SM quark sectors, for completely generic Yukawa interactions there will be
severe constraints on the masses of the quark partners and/or colored gauge bosons $G^\alpha$ requiring that their masses be
$\gtrsim 10$~TeV.  As we shall see below, it is desirable for
these states to appear with much lower masses, and we will typically
assume (as is often done in the Minimal Supersymmetric Standard Model) that some external principle arranges for  appropriate couplings such that
large contributions to flavor-changing neutral currents can be avoided. 

After $SU(4)$ breaking there is a residual $U(1)_\chi$ global symmetry under which the quark partners
all transform with unit charge,
\bea
&&\hspace{-5mm}\tilde{Q}_L \rightarrow e^{i \theta} \tilde{Q}_L \ , \ \  \ 
\tilde{u}_R \rightarrow e^{i \theta} \tilde{u}_R \ , \ \ \ 
\tilde{d}_R \rightarrow e^{i \theta} \tilde{d}_R \ , \nonumber\\
&& \hspace{-5mm}
Q^\prime_R \rightarrow e^{i \theta} Q^\prime_R \ , \ \ \ 
u^\prime_L \rightarrow e^{i \theta} u^\prime_L \ , \ \ \ 
d^\prime_L \rightarrow e^{i \theta} d^\prime_L \ . 
\label{symmetryX}
\eea
This symmetry insures that the lightest of the quark partner fields is stable, and that all of the heavier ones
will decay down to the lightest one.  In that sense it is a continuous analogue to the $R$-parity 
often invoked
in supersymmetric theories of dark matter, though in the present context it is a natural consequence of the
original $SU(4)$ gauge symmetry together with the embedding of the SM quarks into its framework.\footnote{For a model
in which the stability of dark matter can be connected to lepton number, see e.g. \cite{Batell:2011tc,Schwaller:2013hqa}.}

\subsection{Higher Dimensional Operators}

We conclude the discussion of this section by tabulating a few of the most relevant nonrernormalizable interactions
consistent with the gauge symmetries and leading to violations of the global symmetries. 

At dimension five, there are operators involving the $SU(4)$-breaking Higgs which violate $U(1)_\chi$ 
and lepton number, leading to fast decay of the dark matter:
\bea
\hspace{-2mm}
\frac{1}{\Lambda_5} \left[ c_1\,
\bar{\hat{Q}}_L \hat{\Phi} H e_R 
+ c_2 \,\bar{\hat{u}}_R \hat{\Phi} (H \epsilon \,l_L) + c_3 \,\bar{\hat{d}}_R \hat{\Phi} H^* l_L
\right],
\eea
where $\Lambda_5$ characterizes the scale at which the new physics operates and the $c_i$ are complex dimensionless coefficients
describing their relative weighting.
These operators may be forbidden by imposing a $Z_2$ symmetry, and we will not consider them further.

Moving on to dimension six, there is a pair of operators which violate both $U(1)_\chi$ and conventional baryon number:
\bea
\frac{1}{\Lambda_6^2}  \, \epsilon_{abcd}    \left[
c_4\,
\hat{u}^a_{R} \hat{u}^b_{R} \hat{d}^c_{R} \hat{d}^d_{R} + c_5 ( \hat{Q}^a_{L} \epsilon\, \hat{Q}^b_{L})( \hat{Q}^c_{L} \epsilon \,\hat{Q}^d_{L}) 
 \right]
  , \label{dim6ops}
 \eea
which are much more difficult to forbid based on a discrete symmetry.  They ultimately lead to the decay of the dark matter.  Given the
astrophysical bound on the dark matter lifetime of $ \tau_{\rm \,DM}  \gtrsim 4.5 \times 10^{18}  {\rm \ s}$ \cite{Audren:2014bca}, a
viable theory containing dark matter of mass $\sim$~GeV will result provided (assuming the $c_i$ are order one):
 \bea
 \Lambda_{6}  \gtrsim  5 \times 10^{10} {\rm \ GeV}\ .
 \eea
 As we will see below, these operators can also be useful to catalyze an asymmetry among both the baryons and in the dark matter sector.
 
The most relevant operators mediating proton decay are dimension seven and involve the $SU(4)$ Higgs field:
\bea
%\hspace{-1mm}
\hspace{-2mm}\frac{1}{\Lambda_7^3} \,\epsilon_{abcd} \!
\left[ c_6\, 
 \hat{u}^a_{R} \hat{u}^b_{R} \hat{d}^c_{R} \hat{\Phi}^d e_{R} + c_7 ( \hat{Q}^a_{L} \epsilon \,\hat{Q}^b_{L}) (\hat{Q}^c_{L}\epsilon \,l_{L})  \hat{\Phi}^d
 \right]\!.
\eea
The experimental limit on the scale $\Lambda_7$ comes from nucleon decay experiments and is approximately given by:
\bea
\Lambda_{7} \gtrsim 10^{12} \ {\rm GeV} \ .
\eea

\section{Baryogenesis and  Asymmetric \hspace{30mm}Dark Matter} 

The structure of $SU(4)$ enforces a connection between the dark matter number and conventional baryon number.  Indeed, $SU(4)$
implies that at short distances they are precisely the same global symmetry.  However, the dynamics of $SU(4)$ per se does not result in an
asymmetry between particles and anti-particles in either sector.  One option to generate an asymmetry is to invoke the
dimension-six operators of Eq.~(\ref{dim6ops}), which fall out of equilibrium at high temperatures of order $\Lambda_6  \gtrsim  5\times 10^{10} {\rm \ GeV}$ and generically
violate both $C$ and $CP$.

One could construct a specific UV completion of these operators and study the conditions on parameters through which a large enough asymmetry is generated.
However, we choose to remain agnostic about the nature of the UV completion and take for granted that there exist choices of parameters such that
their freeze-out process can generate
a nonzero asymmetry.  Given the high scale at which they act, these choices are unlikely to lead to unique predictions for low energy observables which
would be measurable by current generation experiments.

The structure of the interaction involves three ordinary quarks and one quark partner, and thus the dimension-six interactions create an
initial asymmetry in baryon number, $\Delta B_i$, which balances the asymmetry in the dark matter number, $\Delta \chi$,
\bea
\Delta B_{i} = - \Delta \chi\ .
\eea
The $SU(4)$ interactions, including its instantons, will maintain this relation.  After $SU(4)$ breaking, the most important remnant will be the usual
QCD instantons which preserve equilibrium between chiralities of the light quarks.

As the Universe evolves, the
initial $B$ will be processed into a mixture of $B$ and $L$ by the $SU(2)$ instantons \cite{Harvey:1990qw},
resulting in a modest washout of the baryon asymmetry
(until the electroweak phase transition renders them impotent).
If the masses of the electroweak charged quark partners are somewhat above the electroweak scale, the equilibrium conditions essentially boil down to the
SM ones \cite{Harvey:1990qw}, leading to a final baryon asymmetry:
\bea
\Delta B_{f} = -\frac{28}{79} \,  \Delta \chi\ .
\eea
There will be very small changes (on the order of a percent) to the coefficient if some of the electroweak charged quark partners are lighter than the electroweak scale.

Given this prediction for the relation between the asymmetries in the baryonic and dark sectors (and assuming that the symmetric component of the dark matter
efficiently annihilates away), the mass of the dark matter resulting in the observed ratio of baryonic to dark matter \cite{Ade} can be inferred:
\bea
m_{\tilde{u}'_1} =  \bigg|\frac{\Delta B_{f} }{\Delta \chi}\bigg|\, \frac{\Omega_{\rm DM}}{\Omega_{B}} \, m_{\rm proton} \simeq 1.75 {\rm \ GeV}\ .
\label{DMmass2}
\eea
This mass may be engineered through small ($\lesssim 10^{-3}$) Yukawa couplings for the quark partner playing the role of dark matter.
\footnote{One could also imagine a more standard symmetric dark matter scenario for a Yukawa coupling of order one. 
This would imply a dark matter mass at the $SU(4)$ breaking scale. Although the connection 
to baryogenesis is not clear in this case, it is an option still worth considering since it opens new annihilation channels through intermediate $SU(4)$ and electroweak 
gauge bosons, and such a  model can be probed by direct detection experiments.}

\vspace{-1mm}

\subsection{Annihilation of Symmetric Component}

\vspace{-1mm}

A generic challenge for models of asymmetric dark matter is to realize a large enough annihilation cross section such that
the symmetric component will efficiently annihilate away \cite{Kaplan:2009ag,Graesser:2011wi}.  In the $SU(4)$ model, where the characteristic
scale of the bulk of the new physics is $\gtrsim$~TeV, this typically requires additional ingredients.  

It would be economical to explore the choice of a very small quartic coupling $\lambda_4$, such that the $SU(4)$ Higgs mode is light enough to be produced
in $\tilde{u}'_1$ annihilations.  This option could work, but will require fine-tuning to realize the light $\tilde{u}'_1$  without simultaneously
suppressing the coupling to the Higgs.

 \tikzset{
      my box/.style = {draw, minimum width = 2em, minimum height=1em},
    } 
 \tikzset{
particle/.style={thick,draw=black, postaction={decorate},
    decoration={markings,mark=at position .6 with {\arrow[black]{triangle 45}}}},
particle2/.style={thick,draw=black, postaction={decorate},
    decoration={markings,mark=at position .54 with {\arrow[black]{triangle 45 reversed}}}},
gluon/.style={ultra thick, decorate, draw=black,
    decoration={coil,aspect=0}},
 gluon3/.style={thick, decorate, draw=black,
    decoration={coil,aspect=0}},  
Higgs/.style={thick, dashed, draw=black},
gluon2/.style={thick,decorate, draw=black,
    decoration={coil,amplitude=4pt, segment length=5pt}}
    }          

    \begin{figure}[t!]
     \hspace*{\fill}
     \begin{tikzpicture}[node distance=0cm and 1.47cm]
\coordinate[label=left:$\mathlarger{\mathlarger{{\tilde{u}'_1}}}$] (e1);
\coordinate[below right=of e1] (aux1);
\coordinate[above right=of aux1,label=right:$\mathlarger{\mathlarger{{\psi}}}$] (e2);
\coordinate[below=1.8cm of aux1] (aux2);
\coordinate[below left=of aux2,label=left:$\mathlarger{\mathlarger{\bar{\tilde{u}}'_1}}$] (e3);
\coordinate[below right=of aux2,label=right:$\mathlarger{\mathlarger{\psi}}$] (e4);
\draw[particle] (e1) -- (aux1);
\draw[Higgs] (aux1) -- (e2);
\draw[particle2] (e3) -- (aux2);
\draw[Higgs] (aux2) -- (e4);
\node [label={[label distance=1.6cm]-20:$\mathlarger{\mathlarger{{{\tilde{u}'_1}}}}$}] {};
\draw[particle] (aux1) -- node[label=right:$\mathlarger{\mathlarger{}}$] {} (aux2);
\node[below] at (current bounding box.south){};% {label fig one};
\end{tikzpicture}\hspace*{\fill}
     \begin{tikzpicture}[node distance=0cm and 1.6cm]
\coordinate[label=left:$\mathlarger{\mathlarger{{\tilde{u}'_1}}}$] (e1);
\coordinate[below right=of e1] (aux1);
\coordinate[above right=of aux1,label=right:$\mathlarger{\mathlarger{{\psi}}}$] (e2);
\coordinate[below=1.8cm of aux1] (aux2);
\coordinate[below left=of aux2,label=left:$\mathlarger{\mathlarger{\bar{\tilde{u}}'_1}}$] (e3);
\coordinate[below right=of aux2,label=right:$\mathlarger{\mathlarger{\psi}}$] (e4);
\draw[particle] (e1) -- (aux1);
\draw[Higgs] (aux1) -- (e4);
\draw[particle2] (e3) -- (aux2);
\draw[Higgs] (aux2) -- (e2);
\node [label={[label distance=0.93cm]-32:$\mathlarger{\mathlarger{{{\tilde{u}'_1}}}}$}] {};
\draw[particle] (aux1) -- node[label=left:$\mathlarger{\mathlarger{}}$] {} (aux2);
\node[below] at (current bounding box.south){};% {label fig one};
\end{tikzpicture}\hspace*{\fill}%\begin{tikzpicture}[node distance=0.9cm and 1cm]
%\coordinate[label=left:$\mathlarger{\mathlarger{{{{\tilde{u}}'_1}}}}$] (e1);
%\coordinate[below right=of e1] (aux1);
%\coordinate[above=1cm,right=3.3cm,label=right:$\mathlarger{\mathlarger{{\psi}}}$] (e2);
%\coordinate[right=1.3cm of aux1] (aux2);
%\coordinate[below left=of aux1,label=left:$\mathlarger{\mathlarger{\bar{\tilde{u}}'_1}}$] (e3);
%\coordinate[below right=of aux2,label=right:$\mathlarger{\mathlarger{{{\psi}}}}$] (e4);
%\draw[particle] (e1) -- (aux1);
%\draw[Higgs] (aux2) -- (e2);
%\draw[particle2] (e3) -- (aux1);
%\draw[Higgs] (aux2) -- (e4);
%\draw[Higgs] (aux1) -- node[label=above:$\mathlarger{\mathlarger{\psi}}$] {} (aux2);\node[below] at (current bounding box.south){};% \ \ \color{white} \huge{ h } };% {label fig two};
%\end{tikzpicture}\hspace*{\fill}
\caption{Representative diagram for dark matter annihilation.}
\label{fig:test}
\vspace{-2mm}
\end{figure}
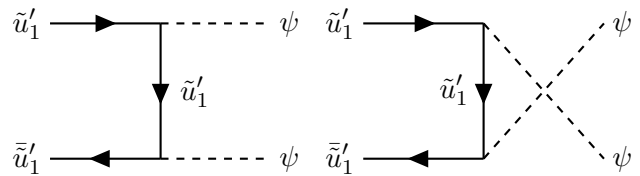   

Perhaps a more appealing option is to introduce an additional light scalar particle $\psi$ which couples to the dark matter,
\bea
 \mathcal{L}_\psi = g \, \psi\, {\bar{\tilde{u}}'_1}{{\tilde{u}}'_1}    \ .
  \label{scalarphi}
 \eea
Such a scalar could arise as the fourth component of an additional $SU(4)$ quadruplet with $X=1/2$, or could be a complete gauge singlet which mixes
with such a state to pick up coupling to the dark matter.  This interaction allows for the annihilation via 
$\tilde{u}'_1 \bar{\tilde{u}}'_1 \rightarrow \psi \,\psi$, provided that $\psi$ is lighter than $\tilde{u}'_1$ (see Fig.~\ref{fig:test}). 
The resulting cross section is velocity suppressed.
Writing the thermally averaged annihilation cross section as
$\langle\sigma_{\!A} v\rangle =    \sigma_0\,T/m_{\tilde{u}'_1}$,
the energy density of relic dark matter particles in the present epoch is \cite{Kolb,Griest:1990kh}:
\bea
\Omega_{\tilde{u}'_1} h^2 \simeq\left( \frac{1.75\times 10^{-10}}{{\rm GeV^2}}\right) \frac{1}{\sigma_0 \sqrt{g_*}} \left(\frac{m_{\tilde{u}'_1}}{T_f} \right)^2,
\label{ann2}
\eea
where $T_f$ is the freeze-out temperature and $g_*$ is the corresponding number of relativistic degrees of freedom. 
For a scalar of mass $m_\psi \approx 1 {\rm \ GeV}$ efficient annihilation of the symmetric component requires a relatively modest coupling:
\bea
g & \gtrsim & 0.05  \ .
\label{relationg}
\eea
Variations are possible, such as introducing a cubic interaction for $\psi$ or lifting the velocity suppression by choosing $\psi$ to be a pseudoscalar.
   
   \vspace{-2mm}
   
\section{LHC Phenomenology}

\begin{figure}[t!]
\begin{tikzpicture}[node distance=1cm and 1cm]
\coordinate[label=left:$\mathlarger{\mathlarger{{q}}}$] (e1);
\coordinate[below right=of e1] (aux1);
\coordinate[above=1cm,right=4cm,label=right:$\mathlarger{\mathlarger{{l}}}$] (e2);
\coordinate[right=2cm of aux1] (aux2);
\coordinate[below left=of aux1,label=left:$\mathlarger{\mathlarger{\bar{{q}}}}$] (e3);
\coordinate[below right=of aux2,label=right:$\mathlarger{\mathlarger{{{\bar{l}}}}}$] (e4);
\draw[particle] (e1) -- (aux1);
\draw[particle] (aux2) -- (e2);
\draw[particle2] (e3) -- (aux1);
\draw[particle2] (aux2) -- (e4);
\node [label={[label distance=1.6cm]-10:$\mathlarger{\mathlarger{{{Z'}}}}$}] {};
\draw[gluon] (aux1) -- node[label=above:$\mathlarger{\mathlarger{}}$] {} (aux2);
\node[below] at (current bounding box.south){};% \ \ \color{white} \huge{ h } };% {label fig two};
\end{tikzpicture}
\caption{Production of $l \bar{l} $  through an $s$-channel $Z'$.}  
\label{fig:zp}
\end{figure}
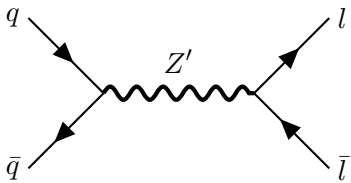

Provided the scale of $SU(4)$ breaking $V$ is low enough, there are signatures related to the additional quark partners and
gauge bosons at the Large Hadron Collider (LHC).  In particular, the existence of an electrically neutral $Z^\prime$ with decays into
leptons furnishes the strongest current bound on $V$.

In the narrow width approximation, the cross section for producing a $Z'$ decaying to $f \bar{f}$
at a hadron collider (see Fig.~\ref{fig:zp}) can be written as \cite{Carena},
\bea
\sigma(p p \rightarrow Z' \rightarrow f \bar{f}) \simeq \frac{\pi}{48 \, s} \sum_{q=u,d} c_q^f \,w_q(s, m_{Z'}^2) \ ,
\eea
where the coefficients $c_q^f$ are given by:
\bea
c_q^f = \left[ (g_q^R)^2 + (g_q^L)^2\right] \, {\rm Br}(Z' \rightarrow f\bar{f}) 
\eea
with $g_q^R$ and $g_q^L$ being the couplings of the $Z'$ to right- and left-handed quarks $q$, respectively, 
and ${\rm Br}(Z' \rightarrow f\bar{f})$ denoting the branching ratio into $f \bar{f}$. 
The functions $w_q(s, m_{Z'}^2)$  contain the parton distribution functions (PDFs) for the proton at the $Z'$ mass scale. 

From Eq.~(\ref{Zcouplings}) the partial $Z'$ decay width into charged leptons is given by:

\vspace{-4mm}

\bea
&&\Gamma(Z' \rightarrow e^+e^-) \simeq \frac{g_Y^2}{24\pi} \left(\frac{5}{4}\tan^2\theta_4\right)m_{Z'} \ .
\eea
The total width of the $Z^\prime$ depends to some degree on the masses of the quark partners.  In the limit where all of the decays are possible, 
the total width becomes:
\bea
\Gamma_{\rm total}(Z') \simeq \frac{g_Y^2}{24\pi}  \left(\frac{4+15\tan^4\theta_4}{\tan^2{\theta_4}}\right)m_{Z'} 
\eea
and the coefficients $c_q^e$ are:
\bea
&& \hspace{-7mm}c_{d/u}^e \simeq  {g_Y^2} \!\left(\frac{2\pm6\,\tan^2\theta_4 +9 \, \tan^4\theta_4}{36\,\tan^2\theta_4}\right) \frac{\Gamma(Z' \rightarrow e^+e^-)}{\Gamma_{\rm total}(Z')} \ . \nonumber\\
%&& \hspace{0mm} = {g_Y^2}\frac{5\tan^2\theta_4\,(2\pm6\,\tan^2\theta_4 +9 \, \tan^4\theta_4)}{72\,(8+21\tan^4\theta_4)}
\eea
Numerically, this corresponds to
\bea
c_d^e &\simeq& 2.1 \times 10^{-4}  \ , \nonumber\\
c_u^e &\simeq& 1.3 \times 10^{-4} \ ,
\eea
for which one can translate the null results from searches for narrow dilepton resonances~\cite{Aad:2014cka,Khachatryan:2014fba}
into a bound on the $Z^\prime$ mass of
\bea
m_{Z'} \gtrsim 2.0 {\rm \ TeV}\ ,
\eea
or in terms of the $SU(4)$ breaking scale,
\bea
V \gtrsim 3.1 {\rm \ TeV} \ .
\label{V}
\eea

\begin{figure}[t!]
\begin{tikzpicture}[node distance=1cm and 1cm]
\coordinate[label=left:$\mathlarger{\mathlarger{{q}}}$] (e1);
\coordinate[below right=of e1] (aux1);
\coordinate[above=1cm,right=3.6cm,label=left:$\mathlarger{\mathlarger{}}$] (e2);
\coordinate[right=1.6cm of aux1] (aux2);
\coordinate[below left=of aux1,label=left:$\mathlarger{\mathlarger{\bar{{q}}}}$] (e3);
\coordinate[below right=of aux2,label= left:$\mathlarger{\mathlarger{{}}}$] (e4);
\coordinate[above right=0.6cm and 2.5cm of e4,label=right:$\mathlarger{\mathlarger{{{\bar{q}}}}}$] (e5);
\coordinate[below right=0.63cm and 1.25cm of e4,label=below left:$\mathlarger{\mathlarger{{{}}}}$] (e5g);
\coordinate[below right=0.6cm and 2.5cm of e2,label=right:$\mathlarger{\mathlarger{{{q}}}}$] (e55);
\coordinate[above right=0.63cm and 1.25cm of e2,label=above left:$\mathlarger{\mathlarger{{{}}}}$] (e55g);
\coordinate[above right=0.4cm and 1.25cm of e5g,label=right:$\mathlarger{\mathlarger{{{\bar{u}}}}}$] (e5g1);
\coordinate[below right=0.45cm and 1.25cm of e5g,label=right:$\mathlarger{\mathlarger{{{\bar{\tilde{u}}}'_1}}}$] (e5g2);
\coordinate[above right=0.45cm and 1.25cm of e55g,label=right:$\mathlarger{\mathlarger{{{\tilde{u}'_1}}}}$] (e55g1);
\coordinate[below right=0.4cm and 1.25cm of e55g,label=right:$\mathlarger{\mathlarger{{{u}}}}$] (e55g2);
\draw[particle] (e1) -- (aux1);
\draw[particle] (aux2) -- (e2);
\draw[particle2] (e3) -- (aux1);
\draw[particle2] (aux2) -- (e4);
\draw[particle2] (e4) -- (e5);
\draw[gluon] (e4) -- (e5g);
\draw[particle] (e2) -- (e55);
\draw[gluon] (e2) -- (e55g);
\draw[particle2] (e5g) -- (e5g1);
\draw[particle2] (e5g) -- (e5g2);
\draw[particle] (e55g) -- (e55g1);
\node [label={[label distance=1.45cm]-11:$\mathlarger{\mathlarger{{{Z'}}}}$}] {};
\node [label={[label distance=3.6cm]6:$\mathlarger{\mathlarger{{{G^\alpha}}}}$}] {};
\node [label={[label distance=4.3cm]-34:$\mathlarger{\mathlarger{{{G^\alpha}}}}$}] {};
\node [label={[label distance=2.6cm]-2:$\mathlarger{\mathlarger{{{{\tilde{q}'}}}}}$}] {};
\node [label={[label distance=3cm]-32:$\mathlarger{\mathlarger{{{\bar{\tilde{q}}'}}}}$}] {};
\draw[particle] (e55g) -- (e55g2);
\draw[gluon] (aux1) -- node[label=above:$\mathlarger{\mathlarger{}}$] {} (aux2);\node[below] at (current bounding box.south){};% \ \ \color{white} \huge{ h } };% {label fig two};
\end{tikzpicture}
\caption{Resonant production of $\tilde{q}' \bar{\tilde{q}}' $ and subsequent decays.}  
\label{fig:qpartner}
\end{figure}
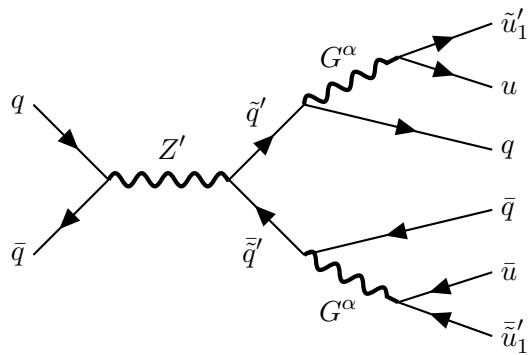

The quark partners have somewhat more model-dependent signatures.  They typically decay through the (off-shell) colored gauge bosons into two quarks
and the dark matter, and thus each one produces a signature typical of a gluino in the Minimal Supersymmetric Standard Model.  However,
they are uncolored and do not have large production cross sections from gluon fusion, and thus will be much more weakly
constrained than gluinos.  Longer cascade decays through
intermediary quark partners are also possible.
The quark partners can be produced via:
\begin{itemize}
\item Drell-Yan-like production because of their electroweak couplings, through an off-shell $\gamma / Z$ or $W^\pm$;
\item together with a quark through the decay of one of the colored gauge bosons; or
\item in pairs through a $Z^\prime$ decay (see Fig.~\ref{fig:qpartner}).
\end{itemize}
These production modes typically produce more kinematic structure than is typical of gluino production (for example, a $Z^\prime$ decay will tend to produce
particles with energy of the order $m_{Z'} /2$).  We leave a detailed recasting of the bounds from searches for jets plus missing momentum for future work.

\section{Conclusions}
We have constructed a novel extension of the Standard Model in which color is unified with baryon number into a single $SU(4)$ gauge group. The theory contains the 
minimal number of new degrees of freedom consistent with the enlarged gauge symmetry. The scale of $SU(4)$ breaking, at which the symmetry of the theory 
reduces to that of the Standard Model, can be as low as a few TeV and all new matter fields are vector-like with respect to the Standard Model gauge group and satisfy 
current LHC constraints.  The structure of the $SU(4)$ demands that the quarks of the Standard Model each come with a partner field that is also a fermion,
uncolored, and with electric charge $\pm 1$ or zero.  The lightest of these states can be neutral (and largely a SM gauge singlet) and is suitable to play the role of dark matter.

The quark partners typically have an unbroken $U(1)$ symmetry which is connected by $SU(4)$ to ordinary baryon number, suggesting that there may be a deep
connection between the apparent stabilities of dark matter and the proton.  It is possible to engineer a situation such that the dark matter is asymmetric, in which case
the lightest quark partner should have a mass around $1.75 {\rm \ GeV}$ and be to good approximation a gauge singlet.  Collider signatures include a rather classic
$Z^\prime$ (though perhaps with exotic decays into the quark partners), and the quark partners themselves, which have a signature similar to gluinos, but are
uncolored and do not have a large pair production cross section, and thus have correspondingly weaker constraints.

\vspace{-3mm}

\subsection*{Acknowledgments}

\vspace{-2mm}
The authors are very grateful to Yuri Shirman for an illuminating discussion about $SU(4)$ instantons, and are
also pleased to acknowledge conversations with Mark Wise.
This research is supported in part by NSF grant PHY-1316792.  TMPT is also pleased to acknowledge support
by the University of California, Irvine through a Chancellor's Fellowship.

\end{document}